\documentclass[11pt,a4paper]{article}

\usepackage[utf8]{inputenc}
\usepackage[T1]{fontenc}
\usepackage{lmodern}
\usepackage[margin=1in]{geometry}
\usepackage{amsmath,amssymb,amsthm}
\usepackage{graphicx}
\usepackage{authblk}
\usepackage{orcidlink}
\usepackage[numbers,sort&compress]{natbib}
\hypersetup{hidelinks}

\graphicspath{{figures/}}
\setcitestyle{authoryear,round}

\title{Autocorrelation-based beam shape optimization in scanning electron microscopy}

\author[1,2]{Rebecca Guan}
\author[3]{Andreas Werbrouck$^{\ast}$\,\orcidlink{0000-0003-3796-0024}}
\author[3,4]{Matthew M. Maschmann$^{\ast}$\,\orcidlink{0000-0001-7384-4333}}
\author[1,3,5]{Matthias J. Young$^{\ast}$\,\orcidlink{0000-0001-7384-4333}}

\affil[1]{Department of Chemical and Biomedical Engineering, University of Missouri, Columbia, MO 65201, USA}
\affil[2]{Department of Chemical and Biomolecular Engineering, Cornell University, Ithaca, NY 14850, USA}
\affil[3]{University of Missouri Materials Science and Engineering Institute, Columbia, MO 65201, USA}
\affil[4]{Department of Mechanical and Aerospace Engineering, University of Missouri, Columbia, MO 65201, USA}
\affil[5]{Department of Chemistry, University of Missouri, Columbia, MO 65201, USA}

\date{}

\begin{document}

\maketitle

\begin{center}
\small
$^{\ast}$Corresponding authors:\\
\href{mailto:andreas.werbrouck@ugent.be}{andreas.werbrouck@ugent.be},\\
\href{mailto:MaschmannM@missouri.edu}{maschmannm@missouri.edu}, or\\
\href{mailto:matthias.young@missouri.edu}{matthias.young@missouri.edu}
\end{center}

\begin{abstract}
Many automatically focusing algorithms for scanning electron microscopes exist. Typically, these optimize a single derived metric (such as image sharpness) without considering the intrinsic beam shape. We describe an algorithm to automatically focus and stigmate scanning electron microscopes based on the autocorrelation of captured images. In the first part of this work, we develop the theoretical basis of using the autocorrelation to approximate the current beam shape. In the second part, we use this to condition the beam shape on various surfaces, using Bayesian optimization to tune both working distance and, in particular, stigmation. 
Through a comparison of the full width at half maximum of the autocorrelation arrays before and after optimization and a visual comparison of images taken of the sample surface before and after optimization, we show that the algorithm is able to successfully optimize the scanning parameters. Overall, our algorithm can effectively focus on a variety of topographies at different magnifications.
\end{abstract}


\section{Introduction}

Scanning electron microscopy (SEM) enables imaging at the micro-and nanoscale through scanning a narrow electron beam across an area of interest. The electrons in the beam interact with the atoms in the sample, generating a signal of varying intensity that can be detected and processed into an image of the sample. Different interactions yield electrons at different energies, giving access to different types of information. Backscattered electrons at high energy are a probe into the composition of the scanned area, whereas secondary electrons, at much lower energy, are typically more sensitive to topography \citep{semelectrons}.

Because the image is constructed through scanning as opposed to lensing (as is the case in light microscopy or transmission electron microscopy), the \textit{shape of the electron beam} determines the quality of the image. Edges in an image will appear sharp if the diameter $d$ of the (circular) beam is smaller than the raster distance $a$. On the other hand, if $d$ is larger than $a$, the beam will act as a convolution kernel for the surface, and the final image will appear blurred.

Especially at high magnifications, it becomes challenging to make the beam sufficiently narrow compared to the scanning grid, and bringing the object(s) of interest within focus (in effect optimizing the beam shape) can take a significant portion of measurement time. Despite technical innovation and decades of autofocus research, this task still depends largely on operator skill and experience. 

Existing plug-and-play solutions, borrowed from lens-based microscopy where the image is formed at once, are not guaranteed to converge to an optimal beam shape. Current autofocus algorithms typically utilize either of two methods. The first method maximizes a sharpness function, such as the image gradient, variance, and Fourier transform, to obtain the parameters corresponding to the optimal beam shape and the most focused image (\citealp{transferfunctions}, \citealp{fft}, \citealp{https://doi.org/10.1002/sca.1998.4950200406}, \citealp{mastersthesis}, \citealp{evaluating}, \citealp{L2normderivative}, \citealp{gradient}, \citealp{bandpassfilters}, \citealp{gaussiandefocus}).

The second method uses machine learning techniques, such as deep learning and neural nets, to score and/or classify images or perform blind deconvolution (\citealp{supportvectormachines}, \citealp{deeplearning}, \citealp{dualdeeplearning}, \citealp{backpropagation}, \citealp{deconvolution}, \citealp{neuralnet}). The advantage of machine learning methods is that they require fewer images of samples to optimize the parameters, thus reducing beam damage to samples. 

What both approaches have in common is that they usually employ a derived metric such as image sharpness (evaluated through various means) to optimize working distance, and potentially also stigmation. The beam shape itself remains a black box. In addition, methods based on a sharpness function are particularly sensitive to noise, requiring high signal-to-noise ratios. Methods based on machine learning require a large amount of microscope-specific (and application-specific) data and training time. They also typically do not generalize well across beam conditions and imaging parameters. 

\begin{figure}[htbp]
    \centering
    \includegraphics[width=0.55\linewidth]{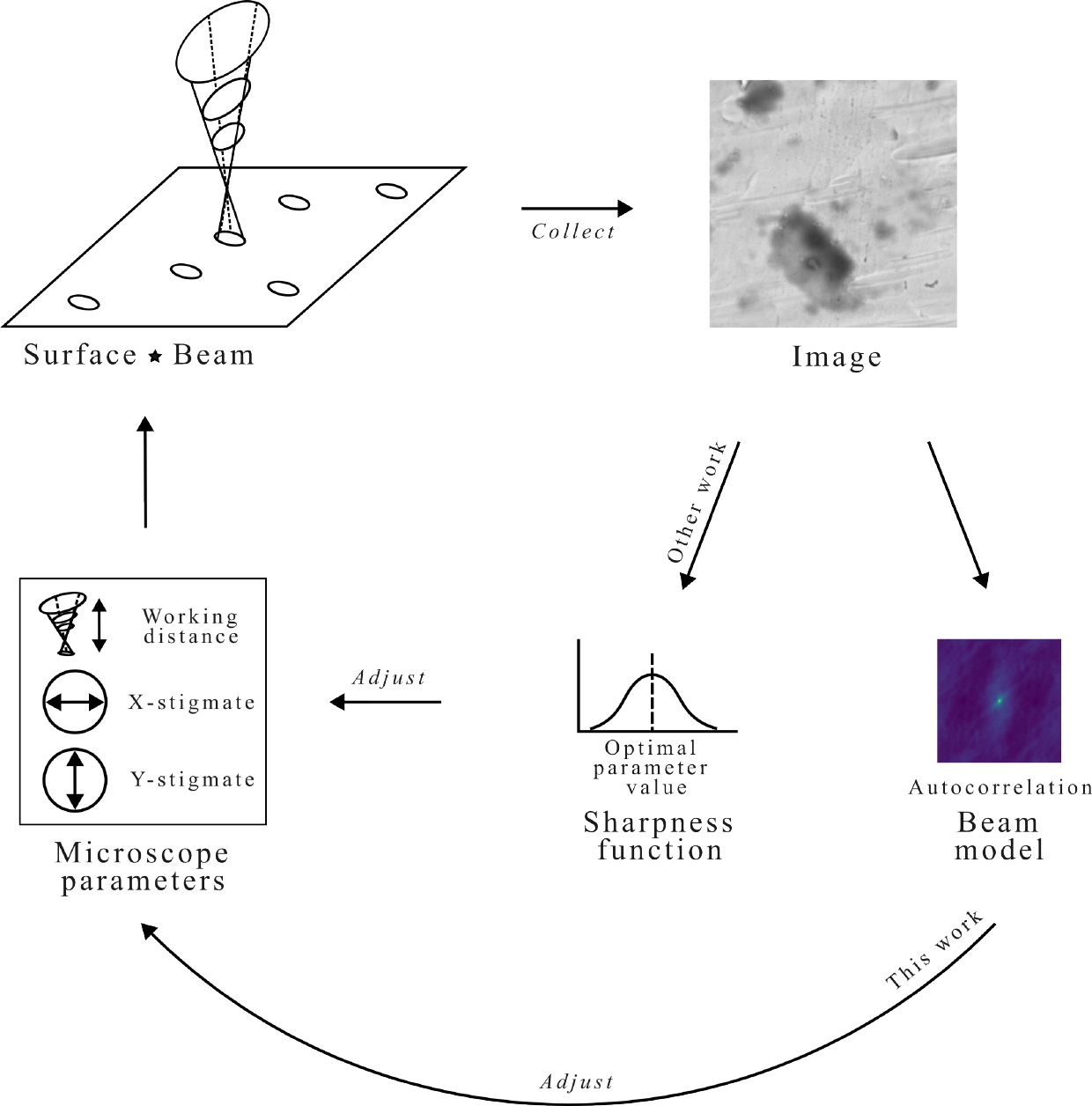}
    \caption{The image formation in a SEM is a convolution between the surface and the shape of the beam. While other methods directly optimize a sharpness function, in this work we attempt to directly model the beam, thus providing a physical interpretation of the optimization of microscope parameters.}
    \label{fig:beamshape}
\end{figure}

Here, we use the autocorrelation of acquired images to estimate the shape of the beam used to scan the surface, and then adjust the microscope parameters to reshape the beam. The workflow is shown in Figure \ref{fig:beamshape}. As developed below, to determine the optimal working distance, a hybrid approach combining edge estimation (a sharpness function) with the beam shape estimation method yields the best results. However, for stigmation, our autocorrelation-based method allows one to directly measure and manipulate the stigmation instead of relying on secondary metrics.

If we define a \textit{ground truth} surface $a(x, y)$, scanned with a beam $b(x,y)$, then we can consider the image $I(x,y) = a (x,y) \star b(x,y) + \epsilon (x,y)$, where the $\star$ symbol denotes a convolution and the $\epsilon$ denotes observational noise. Indeed, in SEM, the point spread function (PSF) $b(x,y)$ is equivalent to the beam shape. Ideally, $b(x,y) = \delta(x,y)$, a two-dimensional Dirac delta function, such that $I(x,y) = a (x,y) + \epsilon (x,y)$, and, in practice, the goal of autofocus algorithms is to bring the beam as close to this ideal shape as possible.

To estimate the state of the beam at a given set of microscope parameters, we employ autocorrelation, defined as the correlation of an image's pixels with shifted versions of itself. Hence, the autocorrelation $R(\xi)$ of a 1D spatial signal $\sigma(x)$ is

\begin{equation}
    R_\sigma(\xi) = \sigma (x)\star \sigma(x + \xi )
    \label{eq:ori}
\end{equation} 
where $\xi$ is the applied shift. Applying this to images, and further developing the notation from the previous paragraph, we get

\begin{align}
    R_I(\xi, \upsilon)  &= (I(x,y) \star I(x + \xi,y + \upsilon)) \nonumber \\
                        &= (a \star b) \star (a \star b) \nonumber \\
                        &= (a \star a) \star (b \star b) \nonumber \\
                        &= R_a (\xi,\upsilon) \star R_b (\xi,\upsilon)
                        \label{eq:autocorr}
\end{align}

using commutative and associative properties of convolutions, with $\xi$ and $\upsilon$, the respective shifts in $x$ and $y$, omitted in the intermediate lines for legibility. This is an important result: the autocorrelation of the image is the convolution of the autocorrelation of the surface itself with the autocorrelation of the beam shape.

In developing this beam shape model, it is useful to examine possible simplifying assumptions in the autocorrelation of $R_a$ and $R_b$. As $R_a$ depends solely on $a(x, y)$, the only definitive general behavior of $R_a$ is that $R_a(0,0) = 1$ will be very sharply peaked in the center, as every pixel perfectly correlates with itself. We do not make any further assumptions on the nature of $R_a$ as we want to be able to focus on any surface $a$ without losing generality. However, it \textit{is} possible to make further assumptions on the beam shape and its autocorrelation by modeling $R_b$ through Zernike polynomials. Another path is to minimize $R_b$ under the assumption that $R_a$ is sharply peaked in the center.

The autocorrelation of an image can be practically calculated through the inverse Fourier transform of the power spectral density. For a discrete 2D image $I(x,y)$, this becomes
\begin{equation}
R_I(\xi, \upsilon) = \mathcal{F}^{-1}\left[\left|\mathcal{F}[I(x,y)]\right|^2\right],
\end{equation}
where $\mathcal{F}$ denotes the Fourier transform, $\mathcal{F}^{-1}$ denotes the inverse Fourier transform, $|\cdot|$ denotes the modulus, and $(\xi, \upsilon)$ are the spatial shifts in the $x$ and $y$ directions. In practice, this is computed using the Fast Fourier Transform (FFT) algorithm and its inverse.

Although the concept of autocorrelation has been used before in other SEM autofocusing algorithms (\citealp{https://doi.org/10.1002/sca.1998.4950200406}, \citealp{mastersthesis}, \citealp{evaluating}), autocorrelation in these works has been used similarly to other sharpness metrics, where the intensity of the autocorrelation signal is correlated to the sharpness of the image. However, in this work, we will develop an algorithm utilizing both autocorrelation as a model for the beam shape and machine learning techniques as an optimization method in the methods section. Afterwards, we will present the results of the algorithm on various substrates and magnifications in the results section.

\section{Methods}

The microscope used was a Phenom Pharos G2 Desktop SEM equipped with a field emission gun (FEG) electron source. Image optimization was achieved through optimization of three parameters: the working distance, $w$, and x- and y- stigmate, $s_x$ and $s_y$. We note that the Phenom system is a tabletop system with a fixed sample height such that the working distance $w$ can only be manipulated through lenses. 

\begin{figure}[htbp]
    \centering
    \includegraphics[width=0.75\linewidth]{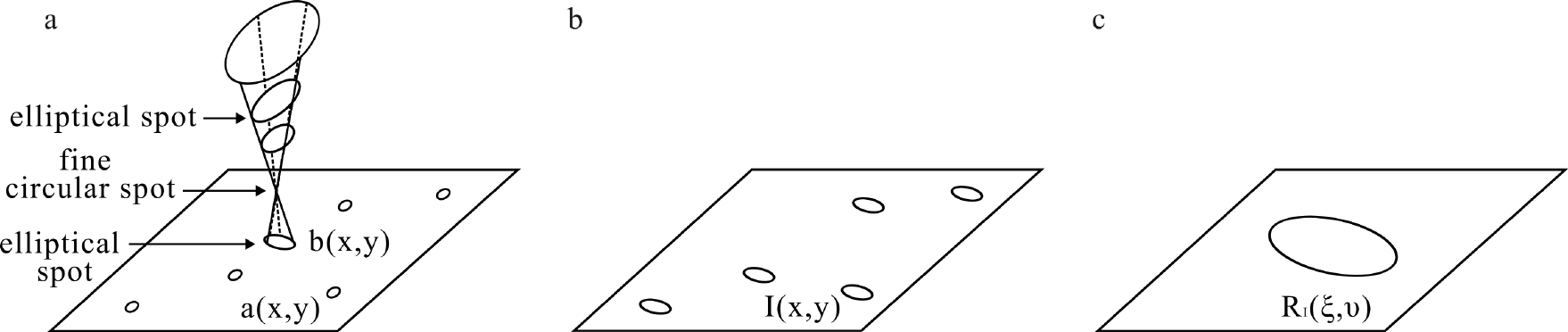}
    \caption{The electron beam cone size and shape, and its convolution with sample features. \textbf{a.} The electron beam shape $b(x,y)$ as it scans across a sample surface with features $a(x,y)$. At a non-optimal working distance, the beam shape is elliptical with a large radius. At the optimal working distance, the beam shape is circular with a small radius. \textbf{b.} The resulting image $I(x,y)$ taken by the SEM using the unfocused beam that is shown in a. As the beam size is larger than the sample feature size, the sample features are convoluted with the beam shape in the image. \textbf{c.} The autocorrelation $R_I(\xi, v)$ of the image in b. The center peak of the autocorrelation has the same shape as the beam shown in a.}
    \label{fig:convolution}
\end{figure}

The working distance $w$, which is the distance between the sample surface and the electron source, determines the beam's focus. It affects the beam radius at the sample surface; as illustrated in Figure \ref{fig:convolution}a, a non-optimal $w$ can result in an under-focused or over-focused state, resulting in the beam being a large, elliptical spot. At the optimal $w$ (focused), the beam is a small circular spot. 

The larger beam radius from an unoptimized $w$ results in a blurry image. Conversely, the smaller beam radius from an optimized $w$ results in a clear image with the pixel values being more distinct from each other. The autocorrelation helps identify image blur because pixels in a blurry image are more correlated to neighboring pixels. An image taken with an optimized $w$ has an autocorrelation peak that is sharper and taller, whereas an image taken with an unoptimized $w$ has an autocorrelation peak that is broader and shorter. 

The x- and y- stigmate ($s_x, s_y$) parameters are corrections of the elliptical character of the beam caused by astigmatism. An unoptimized stigmate $s$ corresponds to the beam having an elliptical shape, resulting in an image with elliptical streaks in place of small features. Conversely, an optimized $s$ corresponds to the beam cone having a circular shape. Autocorrelation is able to identify astigmatism because, under an astigmatic beam condition, pixels will be more similar to each other in \textit{elliptical} regions, whereas under proper stigmation, pixels in a \textit{circular} region will be more similar to each other. In other words, the autocorrelation array from an image obtained with an unoptimized $s$ will have an elliptical peak shape, whereas the autocorrelation array from an image obtained with an optimized $s$ will have a circular peak shape. Additionally, even though an astigmatic beam can be circular at optimal working distance, it can not be as fine as a properly stigmated beam.

This information is summarized in Figure \ref{fig:convolution}. For a sample with features shown in Figure \ref{fig:convolution}a, a beam larger than the sample feature will produce an image where the beam shape and the sample features are convoluted (Figure \ref{fig:convolution}b). As the image represents the convolution of the beam and the sample, the resulting autocorrelation array, specifically the size and shape of the center peak, contains information about the beam size and shape (Figure \ref{fig:convolution}c). 

Consequently, by extracting the information on the beam size and shape contained in the autocorrelation array peak, the set of parameters $w$ and $s$ that correspond to the beam that will result in the sharpest image can be obtained through many methods, such as Bayesian optimization.

\subsection{Working Distance Optimization}

\begin{figure}[htbp]
    \centering
    \includegraphics[width=0.85\linewidth]{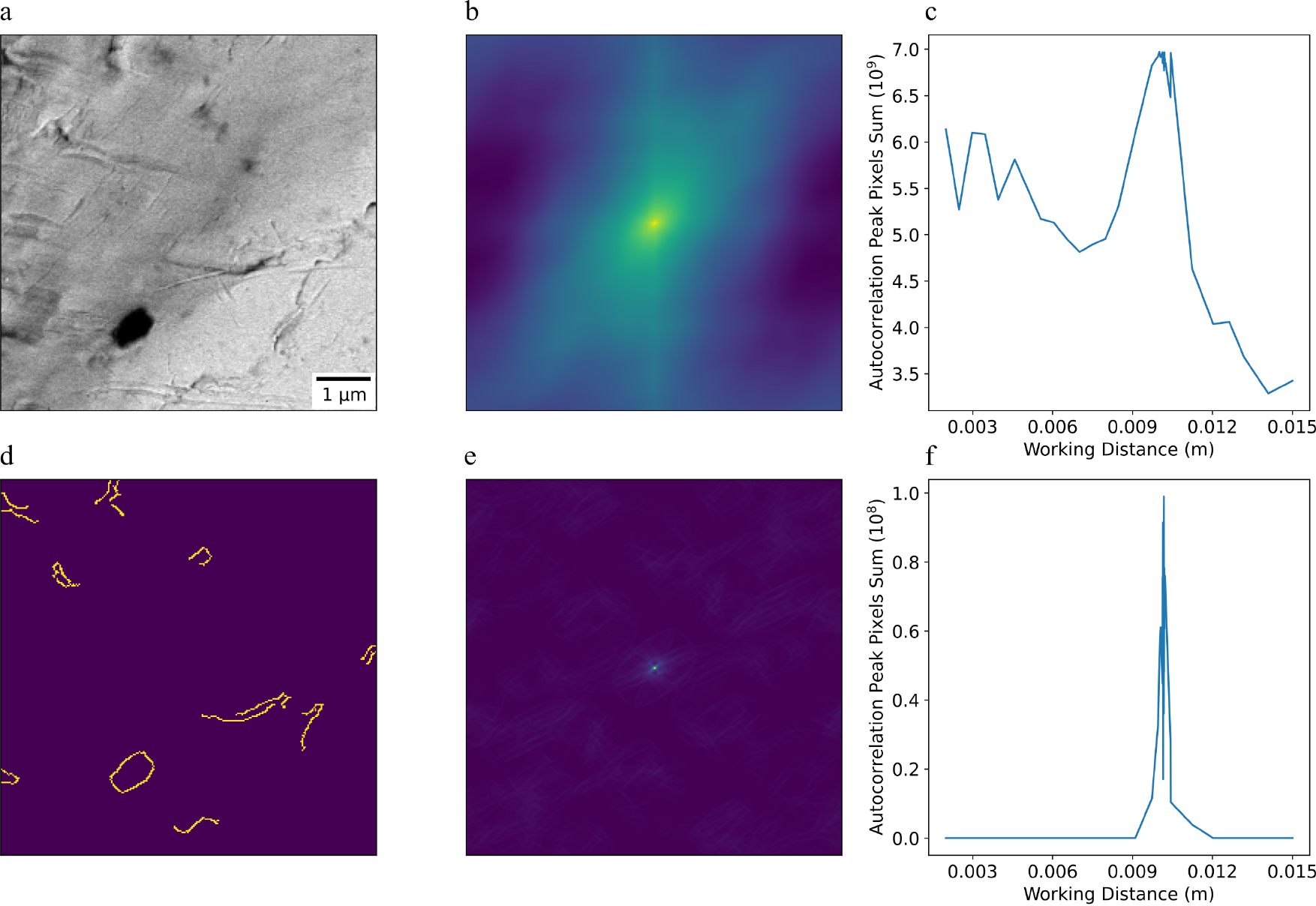}
    \caption{The steps of $w$ optimization. \textbf{a.} The image taken by the SEM. \textbf{b.} The autocorrelation array calculated from the image taken by the SEM. \textbf{c.} The objective function as a function of working distance. The peak is broad, with large fluctuations surrounding it. \textbf{d.} The edges found through a Canny edge detection filter. \textbf{e.} The autocorrelation array calculated from the array with the edges. \textbf{f.} The objective function as a function of working distance. The peak is sharp.}
    \label{fig:wd_figure}
\end{figure}

Figure \ref{fig:wd_figure} shows the different processing steps used in optimizing the working distance using Bayesian optimization. Bayesian optimization was chosen as the optimization method due to its efficiency in optimizing expensive, black-box objective functions, such as the objective functions used for both the working distance and stigmation optimization \citep{bayesianoptimization}. In addition, Bayesian optimization has been used before in aberration-correcting algorithms for electron microscopes, thus supporting the idea that it can be successfully used to automatically focus the SEM (\citealp{physics}, \citealp{neuralnet}, \citealp{beacon}). In this section, we use image autocorrelation in a similar fashion as traditional work uses other sharpness functions.

New values for $w$ are determined through an acquisition function that is probabilistically chosen at each iteration from three models: lower confidence bound, expected improvement, and probability of improvement. These three acquisition functions balance exploitation (testing values where the objective function $f(x)$ is expected to be at a minimum or maximum) and exploration (testing values where the nature of $f(x)$ is uncertain) of the objective function differently.

In the lower confidence bound acquisition function, points are evaluated based on the equation
\begin{equation*}
\alpha(x, \kappa) = \mu(x) - \kappa\sigma(x),
\end{equation*}
where $\mu(x)$ is the predicted mean at $x$, $\sigma(x)$ is the standard deviation at $x$, and $\kappa$ is a trade-off parameter defined by the user (set to 1.96 in this case). In the expected improvement acquisition function, the value expected to most improve the current optimal value of $f(x)$ is evaluated. In the probability of improvement acquisition function, the point most likely to improve the current optimal value of the objective function $f(x)$ is evaluated (\citealp{acquisitionfunctions}, \citealp{sci-kitbayesian}).

The objective function is evaluated over 50 iterations, utilizing 5 or 10 initialization points depending on the phase of the optimization process. Starting from an as-collected 8-bit image taken for each $w$ evaluated (Figure \ref{fig:wd_figure}a), the autocorrelation matrix $R$ with dimensions $M \times N$ (Figure \ref{fig:wd_figure}b) can be calculated. The objective function $f(x)$ being maximized is the sum of the four center values, with a 1 pixel offset from the exact center in each direction, of the autocorrelation matrix, written as:
\begin{equation}
    f(x) = R_{\frac{M}{2} + 1, \frac{N}{2}} + R_{\frac{M}{2} - 1, \frac{N}{2}} + R_{\frac{M}{2}, \frac{N}{2} + 1} + R_{\frac{M}{2}, \frac{N}{2} - 1}.
\end{equation}

This objective function makes the kernel as compact as possible without taking into account long-range correlations in the surface itself. Figure \ref{fig:wd_figure}c shows a systematic evaluation of this function for a large range of working distances. 

This objective landscape, despite having a clear maximum, is still quite rugged. In the interest of increasing performance, we implemented a more traditional Canny edge detection pre-processing step in our data processing pipeline, where the image is smoothed using a Gaussian kernel before applying a Sobel/Canny filter with a lower threshold of 50 and an upper threshold of 150 (Figure \ref{fig:wd_figure}d). The autocorrelation function of Figure \ref{fig:wd_figure}d is then shown in Figure \ref{fig:wd_figure}e.  
This edge detection step helps to eliminate noise, random pixel intensity differences, and false correlations. 

At the beginning of the algorithm, the Gaussian kernel is also optimized to improve performance. At the minimum supportable working distance, an image is taken, and the number of edges in the filtered image is calculated. If this number is not zero, the size of the kernel is increased and the process repeated until no edges are detected. This calibration of the blurring kernel is to ensure a high signal-to-noise ratio so that a strong peak would be obtained at the optimal $w$.

We note that this process introduces additional convolutions, meaning Equation \ref{eq:ori} should be extended to $R_I = R_a \star R_b \star R_c \star R_d$, with $R_d$ being the autocorrelation of the gaussian kernel, and $R_c$ the autocorrelation of the Sobel filter used in the Canny edge detection pipeline. It can be shown that $R_c$ and $R_d$ are centrally peaked. The effect of the Canny edge detection step in the pipeline is a sharper, more defined objective function (Figure \ref{fig:wd_figure}f). Thus, the working distance optimization does not directly optimize the beam shape; rather, it leans toward a more traditional sharpness function-based autofocusing method, with the sharpness function defined by the autocorrelation of the processed image.

\subsection{Stigmate Optimization}
After optimizing the working distance, an astigmatism correction is carried out, using explicit beam modeling, showing the true potential of the autocorrelation method. The method developed in this section is the primary contribution of this work. 

Initially, an image of the current location is taken with the optimized $w$ and the unoptimized $s$ parameters, and the autocorrelation of the image array is calculated. To eliminate background noise in the autocorrelation array arising from topographical similarities rather than beam geometry, the data is first processed. The relevant part of the autocorrelation array is the center peak; however, the background region of the array may be a constant value, waves of smaller peaks, or other textures, such as the image depicted in Figure \ref{fig:wd_figure}b.

The background is removed by determining the locus around the center peak where the second derivative of the autocorrelation array is approximately zero. This boundary represents the end of the center peak and the start of the background. This condition for masking was selected due to the behavior of the arrays, as a change in curvature occurs when the center peak transitions to the background. Thus, using the locus where the second derivative is equal to zero, keeping only the set of points inside the locus for further processing and discarding everything outside, was determined to be a good condition for masking. We note that a range of other background selection criteria were examined, but this criterion provided the best performance among those attempted.

After masking the part of $R_I$ where the background contribution of the autocorrelation is dominant, we consider the autocorrelation $R_b(\xi, \upsilon) = b\star b$. We can model $b$ using the pupil function 
\begin{equation}
    P = A(\rho, \theta) \exp(i \Phi(\rho, \theta))
\end{equation}
where 
\begin{equation*}
\Phi(\rho, \theta) = \sum_{n} \sum_{m} c_n^m Z_n^m(\rho, \theta)
\end{equation*}
and $Z_n^m$ are the Zernike polynomials. 

The pupil function can then be transformed into the PSF $b = |\mathcal{F} [ P(\rho, \theta) ]|^2$. From this, we can calculate $b\star b$ for various prefactors Zernike polynomials, and fit this to the data using a non-linear least-squares regression utilizing the LMFIT library.

By changing the prefactors $c_n^m$, the pupil function can be altered in a physically interpretable way. In a Zernike polynomial, the defocus coefficient ($c^0_2$) models the radius of the shape, and the two astigmatism coefficients ($c^{-2}_2, c^2_2$) model the eccentricity and the direction (horizontal or vertical) of the elliptical stretch. These correspond in a one-to-one fashion to the SEM controls.

As an ideal beam shape maps to a minimized circular center autocorrelation array peak, Bayesian optimization is used to minimize the magnitude of the fitted astigmatism coefficients. The objective function is evaluated over a user defined number of iterations (here 50), utilizing 10 initialization points. Again, the acquisition function used is probabilistically chosen at each iteration from three models: lower confidence bound, expected improvement, and probability of improvement. The objective function, $f(x)$, being minimized is the magnitude of the astigmatism coefficients:
\begin{equation}
    f(x) = \sqrt{(c^{-2}_2)^2 + (c^2_2)^2}.
\end{equation}
The defocus coefficient $c^0_2$ is not considered in the objective function, as it is possible to have a minimized radial value but a highly eccentric shape.

\subsection{Overall Algorithm}
The overall algorithm starts by optimizing $w$ at a large field width (FW) of 104 $\mu$m, ensuring that there is a feature within the field to focus on during the initial optimization. This initial phase consists of a user defined number of Bayesian iterations (here 50), enacting the process explained above, to determine an approximate $w$ parameter. 

Following the initial optimization of $w$, for beam shape estimation, \textit{a} feature is necessary, not necessarily circular. To find such a feature in the field of view, the local entropy of every pixel in the image is used. It is calculated with the equation
\begin{equation*}
H = -\sum_i p_i \log_2 p_i,
\end{equation*}
where $i$ is a pixel intensity that occurs with probability $p_i$, given the histogram of the surrounding area \citep{entropy}. The entropy quantifies the information content and complexity of a given neighborhood in an image and is calculated through the skimage entropy function.

Since a more in-focus image has higher information content, entropy has been used previously as a measure of image sharpness in autofocusing algorithms \citep{https://doi.org/10.1002/cyto.990120302}. In addition, the local entropy can detect variations in pixel intensity in the surrounding neighborhood, which correspond to well-defined features on the sample surface. The region of the field with the largest entropy thus has the largest and most distinct features; in its current implementation, the algorithm automatically repositions on this feature, thus ensuring that a feature will remain in view to aid with focusing even when the FW is decreased. This repositioning is optional if the high entropy feature remains in the field of view.

In addition, the search boundaries for $w$ are changed to the current optimal $w$ $\pm$ 1 millimeter. This narrowing of the sample space is done so that further optimization can probe more points near the current optimal point, thus providing a more accurate determination of the optimum.

The user then defines the target imaging FW. Using the repositioned region and the new $w$ probing range, an additional user-defined number of Bayesian optimization iterations are performed (here 50), thus determining an accurate $w$ at the desired FW.

After the $w$ optimization is complete, the location and FW to correct for astigmatism is determined by hand. The optimal location should be a location with small features relative to the total field. This ensures that the sample features do not dominate the autocorrelation array; if the convolution is mostly the topography, then the algorithm will incorrectly optimize for feature circularity, not beam circularity. 

Following the determination of the location, $s$ is optimized through a user-defined number of Bayesian optimization iterations (here 30) in the process explained above. 

\subsection{Parameter Choice}

Experimental trials determined that the smallest successful working distance $w$ optimization occurred within an FW range of 2.16 $\mu$m to 6.50 $\mu$m, with a smaller FW being obtained in locations where a small, defined feature at that FW was available to be focused on.

The FW range in which we tested stigmation $s$ optimization was 2.53 $\mu$m (10 nm/pixel) to 8.21 $\mu$m (32 nm/pixel). The smaller FWs were obtained from samples where the location in view had small features relative to the total sample field. Due to this condition, the range of FW values for the $s$ optimization range is higher than the $w$ optimization range, as the field width often needs to be increased to find a sample location that satisfies this condition.

Shown below are some highlighted results of the algorithm. To quantify the optimization performance, the autocorrelation arrays before and after optimization are compared through the full width at half maximum (FWHM) of the central peak. The FWHM of the x- and y-direction cross-sections through the center of the autocorrelation array are evaluated using the SciPy signal processing package. The overall FWHM is determined through the equation:
\begin{equation}
    \text{FWHM}_\text{overall} = \sqrt{\text{FWHM}_x^2 + \text{FWHM}_y^2},
\end{equation}
where $\text{FWHM}_x$, the FWHM in the x-direction, and $\text{FWHM}_y$, the FWHM in the y-direction, are added in quadrature. A successful optimization corresponds to a minimization of $\text{FWHM}_\text{overall}$.

\section{Results}

The performance of the algorithm was evaluated on samples with different surface topographies and sample conductivities: a coin with high conductivity and complex, multi-scale features (scratches), and a non-conductive polymer thin film with 120 nm nanoparticles added through spin-coating. The average runtime was 166 seconds. The workflow described here should be considered as a proof of principle, and further optimizations are definitely possible.

\subsection{Coin Surface: Linear Columns of Small Circular Features}

The algorithm was applied to a region of the coin that featured small circular elements arranged in vertical lines, with some larger circular elements scattered randomly. 

\begin{figure}[htbp]
    \centering
    \includegraphics[width=\linewidth]{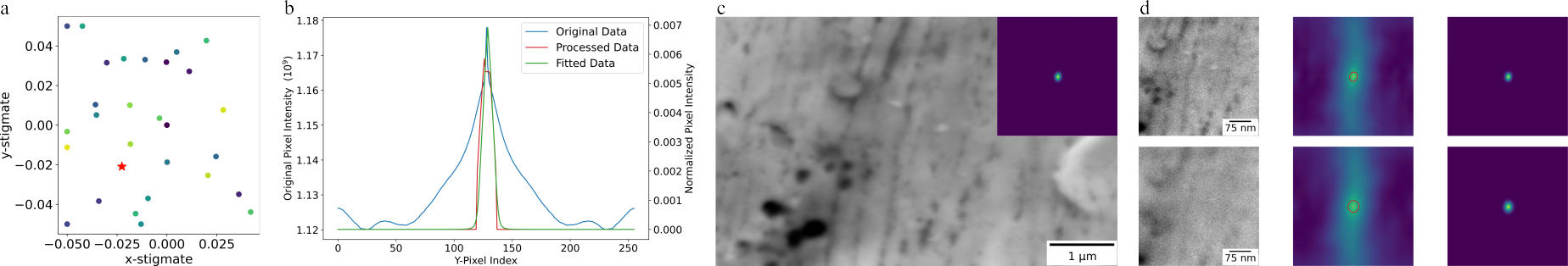}
    \caption{The image taken after optimization and the steps of $s$ optimization. \textbf{a.} The $s_x$ and $s_y$ values sampled when determining the optimum $s$. The darkest color corresponds to the first sampled parameter, and the lightest color corresponds to the last sampled parameter. The optimal $s$ is marked with a red star. \textbf{b.} The original pixel intensity, data pixel intensity after background removal, and fitted data pixel intensity of a cross-section of the autocorrelation array. \textbf{c.} The final image taken at a field width of 6.22 $\mu$m. The final image autocorrelation, showing the beam size and shape, is shown in the top right corner. \textbf{d.} A comparison between the image (first column), autocorrelation (second column), and Zernike fit of the beam shape (third column) for an image taken with an unoptimized $s$ (bottom row) and an optimized $s$ (top row). The red ellipses on the autocorrelation arrays represent the calculated boundary between the center peak and the background values.}
    \label{fig:83stig}
\end{figure}

The final results after optimization at a field width of 6.22 $\mu$m and some of the optimization steps are summarized in Figure \ref{fig:83stig}. 

To determine the optimal $s$, 30 sets of parameters were tested as determined by Bayesian optimization. The stigmation space and the tested parameters are shown in Figure \ref{fig:83stig}a, with the darkest color corresponding to the first sampled parameter and the lightest color corresponding to the last sampled parameter. The optimal $s$ parameter is marked with a red star. The continued spread of the points with increasing number of samples in \ref{fig:83stig}a and similar plots below arises from the stochastic selection of exploration and exploitation criteria implemented in the Bayesian optimization algorithm.

Figure \ref{fig:83stig}b shows a comparison of the cross-sections of the original autocorrelation array, the autocorrelation array after data processing, and the Zernike polynomial fit of the processed data to compare the fidelity of the data processing and fitting routine. As can be seen, the processed data maintains the location and approximate shape of the original autocorrelation array peak, but without the background noise, allowing the fit to accurately approximate only the center peak.

The post-optimization image and its autocorrelation array are shown in Figure \ref{fig:83stig}c. A full comparison between the optimized state and an initial unoptimized state is provided in Figure \ref{fig:83stig}d. 

The unoptimized $s$ autocorrelation, sampled near the boundaries of the sampled range, had a $\text{FWHM}_{\text{overall}}$ of 20.5 pixels, while the optimized $s$ autocorrelation had a $\text{FWHM}_{\text{overall}}$ of 16.1 pixels. Additionally, the optimized image is visually more in-focus than the non-optimized image. These results agree with the prediction that a sharper autocorrelation peak corresponds to a more in-focus image, supporting the autocorrelation array approach used to optimize the parameters.

\subsection{Polymer Thin Film with Nanoparticles}

\begin{figure}[htbp]
    \centering
    \includegraphics[width=\linewidth]{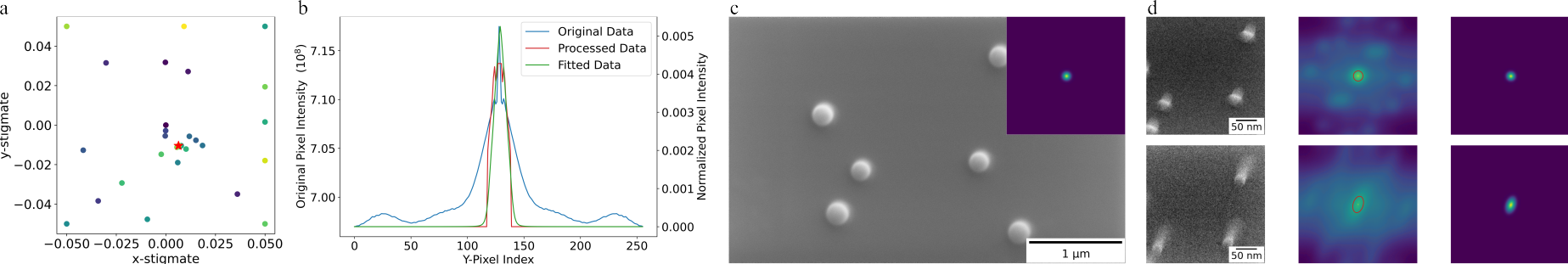}
    \caption{The image taken after optimization and the steps of $s$ optimization. \textbf{a.} The $s_x$ and $s_y$ values sampled when determining the optimum $s$. The darkest color point is the first tested parameter, while the lightest color point is the last tested parameter. The optimal $s$ is marked with a red star. \textbf{b.} The original pixel intensity, data pixel intensity after background removal, and fitted data pixel intensity of a cross-section of the autocorrelation array. \textbf{c.} The final image taken at a field width of 4.27 $\mu$m. The final image autocorrelation, showing the beam size and shape, is shown in the top right corner. \textbf{d.} A comparison between the image (first column), autocorrelation (second column), and Zernike fit of the beam shape (third column) for an image taken with an unoptimized $s$ (bottom row) and an optimized $s$ (top row). The boundary for the background removal is shown as a red line on the autocorrelation arrays.}
    \label{fig:120stig}
\end{figure}

The algorithm was subsequently evaluated on a low-conductivity polymer thin film containing spin-coated 120 nm nanoparticles, operating at a FW of 4.27 $\mu$m (Figure \ref{fig:120stig}).  

The set of parameters in the sample space across the 30 iterations of Bayesian optimization is plotted in Figure \ref{fig:120stig}a. Again, the optimal $s$ parameter is marked with a red star.

Once again, Figure \ref{fig:120stig}b compares the cross-sectional profiles of the autocorrelation arrays before and after data processing to verify the validity of the data processing approach.

Finally, the resulting optimized image of the sample surface is shown in Figure \ref{fig:120stig}c, with the corresponding autocorrelation array shown in the top right corner. 

A full comparison between the unoptimized and optimized $s$ states is shown in Figure \ref{fig:120stig}d. The minimization of the FWHM, from $\text{FWHM}_{\text{overall, unopt}} = 24.6$ pixels to $\text{FWHM}_{\text{overall, opt}} = 17.8$ pixels, with the larger FWHM corresponding to the visually blurrier image, again demonstrates that a minimized autocorrelation array peak corresponds to a focused image. This result also establishes the capability of this algorithm to optimize the beam parameters on a non-conducting substrate.

\subsection{Coin Surface: Large Circular Features}

\begin{figure}[htbp]
    \centering
    \includegraphics[width=\linewidth]{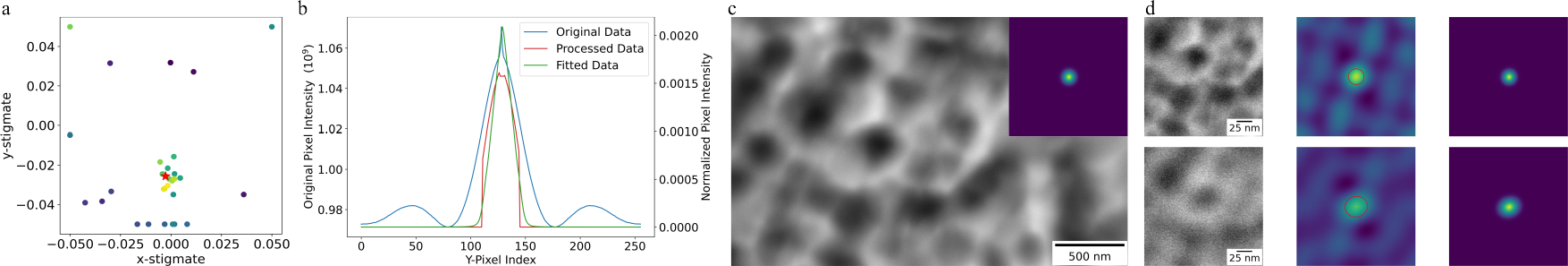}
    \caption{The image taken after optimization and the steps of $s$ optimization. \textbf{a.} The $s_x$ and $s_y$ values sampled when determining the optimum $s$. The darkest color point represents the first tested parameter, while the lightest color point represents the last tested parameter. The optimal $s$ is marked with a star. \textbf{b.} The original pixel intensity, data pixel intensity after background removal, and fitted data pixel intensity of a cross-section of the autocorrelation array. \textbf{c.} The final image taken at a field width of 2.86 $\mu$m. The final image autocorrelation, showing the beam size and shape, is shown in the top right corner. \textbf{d.} A comparison between the image (first column), autocorrelation (second column), and Zernike fit of the beam shape (third column) for an image taken with an unoptimized $s$ (bottom row) and an optimized $s$ (top row). The boundary for the background removal, separating the center peak and the background correlations, is shown as a red ellipse on the autocorrelation arrays.}
    \label{fig:180stig}
\end{figure}

Finally, the algorithm was applied to a region of the coin featuring dense circular features separated by narrow gaps, optimized at a FW of 2.86 $\mu$m (Figure \ref{fig:180stig}). 

Again, the sampling space and the confirmation of data processing fidelity are shown in Figures \ref{fig:180stig}a and \ref{fig:180stig}b.

Figure \ref{fig:180stig}c contains the final image after optimization with the corresponding autocorrelation shown in the top right corner. Comparing the initial and final states and their autocorrelation arrays (Figure \ref{fig:180stig}d) shows that the unoptimized $s$ parameter led to a blurry image with a $\text{FWHM}_{\text{overall}}$ of 38.7 pixels, while the optimized $s$ parameter resulted in a more clear image with a $\text{FWHM}_{\text{overall}}$ of 28.6 pixels. This further validates the autocorrelation optimization approach used to obtain these results.

\section{Discussion}
Although these trials demonstrated the effectiveness of the algorithm on a variety of substrates, they also highlighted a limitation of the stigmation optimization process. When the sample surface's autocorrelation dominates the overall autocorrelation array, then the optimization process attempts to optimize the sample shape to be spherical, rather than the beam shape. This situation happens most frequently when there are large topographical features, relative to the field size, and can thus be avoided by finding a region of the sample where the features are relatively small, or changing the FW so that the features are smaller relative to the total field. In this respect, we refer back to Equation \ref{eq:autocorr}: our entire workflow assumes that $R_b \gg R_a$, which is violated when large features across the entire field of view are present.

We also point out that our approach relies on $b$ being symmetric and real-valued. Indeed, this implies that the autocorrelation and the convolution are identical. As such, our approach only works for symmetric beam aberrations.

Furthermore, because the algorithm was tested on a tabletop SEM, there are intrinsic limitations on the quality of focus obtainable, as the SEM does not have the instrumentation to provide higher-level corrections when focusing. Consequently, adapting this algorithm to higher-resolution microscopes would require adjusting the optimization parameters to allow for higher-order corrections.

One can easily imagine that a direct estimation of beam shape from collected images would be of interest to correct for other microscope parameters as well. The probe current and accelerating voltage will also have a large impact on the quality of the image, especially at high magnifications. Another relevant parameter could be deceleration voltage at the stage, should the microscope permit, to allow for better imaging of isolating samples.

\section{Conclusions}

To facilitate the use of scanning electron microscopes, robust and efficient autofocusing algorithms are needed. In this paper, we developed an autofocusing algorithm based on the autocorrelation of images taken by the SEM. In this method, the autocorrelation estimates the beam shape itself, resulting in the optimization of the scanning parameters through the optimization of the beam shape. This is in contrast to more traditional approaches, where the scanning parameters are optimized through the metric of a sharpness function. In addition, we test the algorithm on a variety of topographies and magnifications, and we conclude that this algorithm is a promising method for automatically focusing scanning electron microscopes. 

\section*{Competing interests}
No competing interest is declared.

\section*{Acknowledgments}
R.G. and M.J.Y. acknowledge partial support from the National Science Foundation Division of Materials Research (DMR) Polymers Program through Award Number 2235161. A.W. acknowledges support from the University of Missouri Materials Science and Engineering Postdoctoral Fellowship. M.J.Y. acknowledges partial support from the Division of Graduate Education NRT Program through Award Number 2243526.

\bibliographystyle{apalike}
\bibliography{ref}

\end{document}